\documentclass[aps,prc,twocolumn,floatfix,showpacs,a4paper,
nofootinbib,amsmath,amssymb]{revtex4-1}
\usepackage{graphicx}
\usepackage{dcolumn}
\usepackage{bm}
\usepackage{color}

\newcommand{\be}{\begin{equation}}
\newcommand{\ee}{\end{equation}}
\newcommand{\ba}{\begin{eqnarray}}
\newcommand{\ea}{\end{eqnarray}}
\newcommand{\bd}{\begin{displaymath}}
\newcommand{\ed}{\end{displaymath}}
\renewcommand{\vec}[1]{\mbox{\boldmath$#1$}}

\begin{document}
\title{Viscous potential flow analysis of peripheral heavy ion collisions}

\author{D.J. Wang$^1$, Z. N\'eda$^2$, and L.P. Csernai$^{1}$}

\affiliation{
$^1$
Department of Physics and Technology,
University of Bergen, Allegaten 55, 5007 Bergen, Norway \\
$^2$
Department of Physics, Babe\c{s}-Bolyai University, Cluj, Romania
}

\begin{abstract}
The conditions for the development of a Kelvin-Helmholtz
Instability (KHI) for the Quark-gluon Plasma (QGP) flow
in a peripheral heavy-ion collision is investigated.
The projectile and target side particles are
separated by an energetically motivated
hypothetical surface, characterized with a phenomenological
surface tension. In such a view, a classical potential flow
approximation is considered and the onset of the KHI is studied.
The growth rate of the instability is computed as
function of phenomenological parameters characteristic for the
QGP fluid: viscosity, surface tension and flow layer thickness.
\end{abstract}

\date{\today}


\maketitle

\section{Introduction}

The first models of high energy heavy ion collisions
in the 1970s were successful by assuming highly
idealized shock fronts where the matter was heated up
and compressed in a front (having a discontinuity in
perfect fluid flow
\cite{Greiner1974,Chapline1973}).
This led to high pressured, shock compressed
domains which collectively deflected the incoming
nuclear fluid. The observation of this directed
flow (side splash of bounce-off) was the first
proof of the collective fluid dynamical behaviour
of nuclear matter \cite{Gustafsson1984}. Recent theoretical
developments and experimental observation of high
multipolarity fluctuations indicate that the QGP is a low
viscosity fluid, which makes turbulent phenomena possible
\cite{CSA11,Bonasera,Stefan}

\begin{figure}[ht]  
\begin{center}
\resizebox{0.9\columnwidth}{!}
{\includegraphics{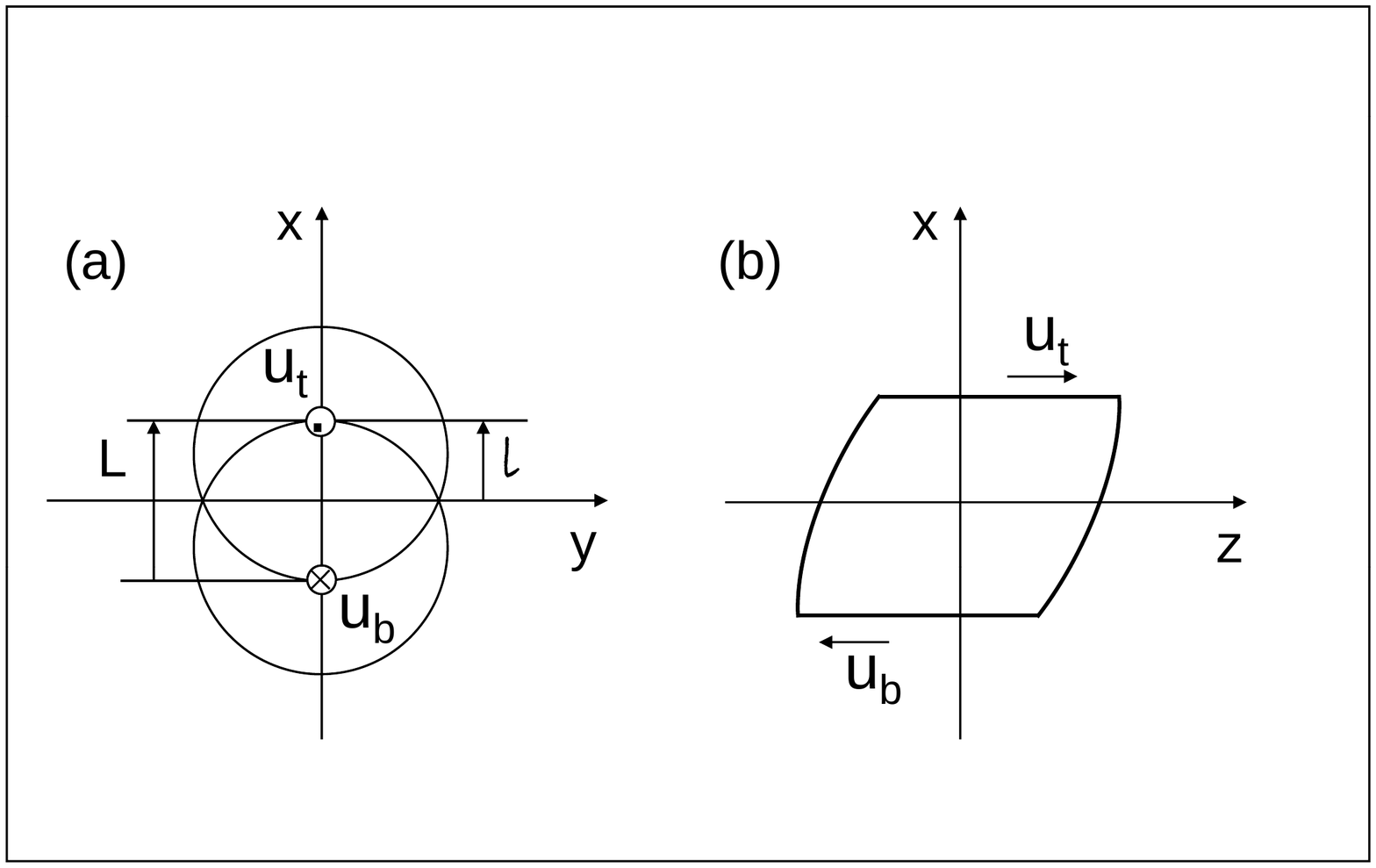}}
\caption{
Sketch of a collision: (a) is a view in the transverse, [x,y] plane and
(b) is an illustration in the reaction, [x,z] plane. The almond
shape in the middle of figure (a) is
the participant zone of the event. Right after the collision, streaks are
formed and the top streaks move along the $z$ direction while bottom ones
move along the $-z$ direction. Due to this velocity shear, an
instability wave will
appear on the interface plane between the top and bottom sheets.
}
\label{sketch}
\end{center}
\end{figure}

Here we adopt again a fluid dynamical picture and discuss
the strong shear flow arising in the
initial states of peripheral heavy ion collisions
at ultra-relativistic energies, which may lead to
KHI under favorable conditions, as discussed recently
\cite{CSA11}.

A simple analytic study showing the development of the KHI
in a highly idealized situation is discussed. The investigated phenomenon has
some resemblance to the initial state of a peripheral heavy ion collision.
In these collisions the collective flow should be a "shear flow" because
the top participant layers, move nearly with projectile velocity while the
bottom layers with the target velocity.

In the reaction plane, the height of the participant profile is $L=2R-b$,
where $b$ is the impact parameter, and the half
height is $l=L/2=(2R-b)/2$. In the following we will denote by $n$ the
nuclear matter density and by $\eta$ its phenomenological viscosity.
The matter coming from
different sources is marked by subscript/superscript $t$ (top) and $b$ (bottom),
see Fig. \ref{sketch}.

For simplicity reasons we consider a two dimensional
non-relativistic dynamics in the reaction plane.
The position vector is $\vec{x}=(x,z)$,
and the velocity vector is $\vec{v}=(v_x,v_z)$.

\vskip -0.4cm
\begin{figure}[ht]  
\begin{center}
\resizebox{0.55\columnwidth}{!}
{\includegraphics{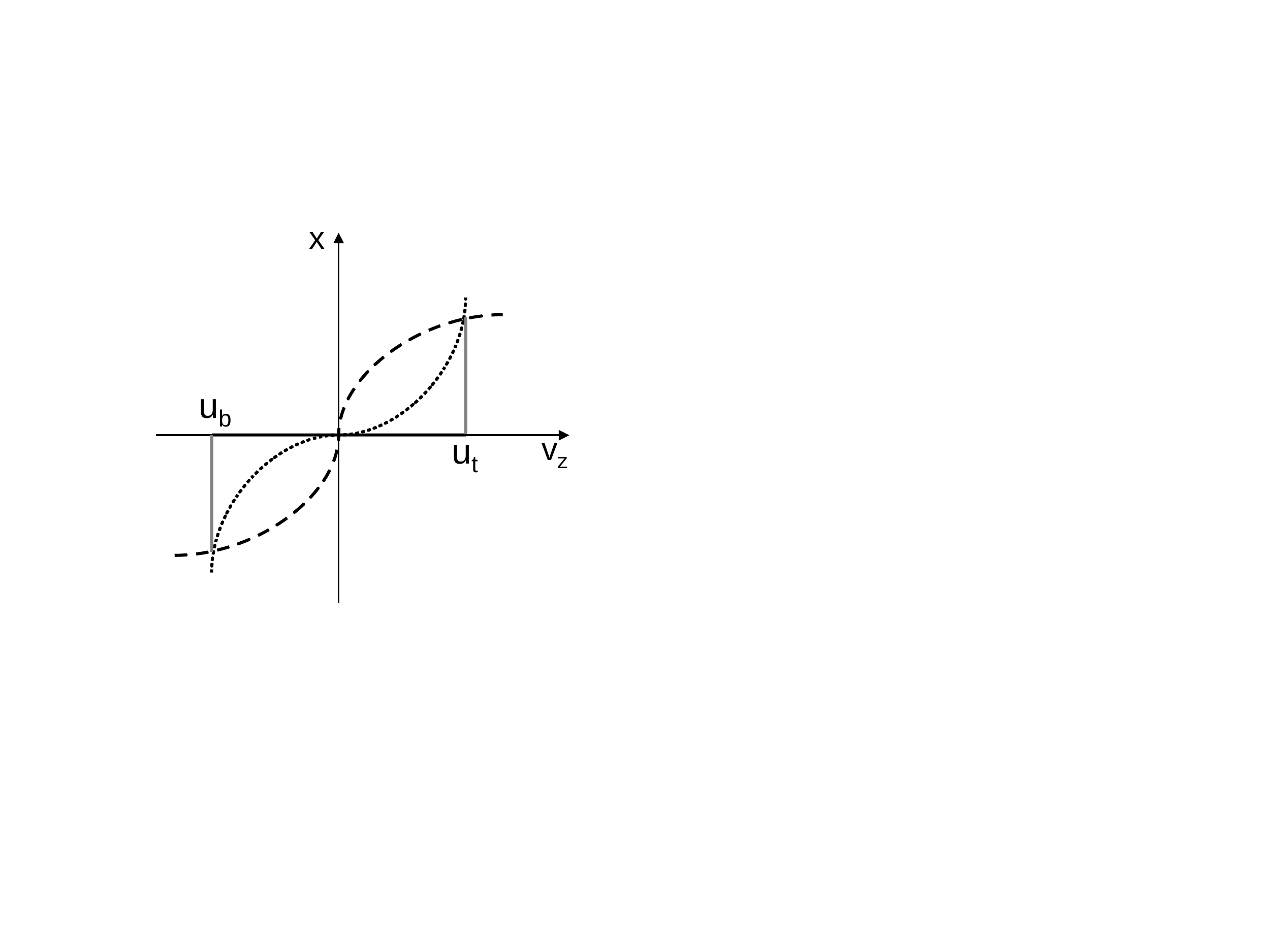}}
\caption{
The velocity along the $z$ axis, is represented by the dotted curve,
calculated in our CFD model and presented in Ref. \cite{CSA11}.
The full line
with two singularities in its derivative is the case used in
Ref. \cite{FUNADA}.
Here we have to mention that the velocity profile of the dotted
curve will induce
the KHI effect, while the velocity profile illustrated with the
dashed curve will not, see chapter 8
of Ref. \cite{Drazin}.
}
\label{sketch3}
\end{center}
\end{figure}
\vskip -0.4cm

The nearly perfect QGP
provides a possibility of a strong idealization in this
situation. The velocity profiles for $v_z$ presented on
Fig. \ref{sketch3} illustrates
(the dotted line) that the KHI develops
if we have a strong shear flow at the
$x=0$ plane, which leads to large vorticity and
circulation. With decreasing viscosity we could
idealize this configuration in a way that the
vorticity is constrained into a narrow layer around
the $x=0$ plane, while the circulation remains constant.
In a limiting case if we constrain the vorticity and shear
to the dividing, $x=0$, plane, this plane would represent
an infinite vorticity, providing the same circulation
for a trajectory surrounding the dividing plane.
\footnote{The conservation of circulation occurs in
classical, barotropic flow. In QGP the temperature
dominates the pressure change, so the circulation is
not conserved but decreases during the expansion
of the system \cite{Vorticity}.}
In this limiting case the velocity profile is
idealized to the one indicated by the full line in Fig. \ref{sketch3}.
Thus in the top and bottom domain the flow has no shear,
and can be described as potential flow, which is an
important simplification and idealization. This makes the
analytic study of the KHI possible.

\section{The analytic model}

Thus, the idealized dividing layer represents a discontinuity
of the flow velocity (i.e. unconstrained slip). At the same time a
small viscosity would contribute the transverse momentum
transfer to a small distance across the dividing front. The
particles in this narrow layer, scattering over from the other side of the
dividing plane, would have a high relative velocity, so this
layer would exhibit an extra energy increase compared to the
general fluid body on the top or the bottom side of our system.
This can be taken into account as an effective surface energy
of the dividing layer. This surface energy can be estimated both
from a microscopic kinetic theory approach, or from a rough energy
balance calculation.  In a microscopic approach one would assume that the
this extra energy depends on the (viscosity dependent)
thickness of the layer and the (temperature dependent) rate of
transverse flow crossing the dividing plane. A quantitative estimate
of this surface energy in this idealized situation is not feasible,
but its existence and a rather qualitative estimate can be made.
In contrast with this, a phenomenological energy balance
calculation is easier
to perform. The advantage of such an approach would be
that one does not rely on further estimates
for the involved physical parameters.
Here, we will use this later method to approximate the surface tension of the
dividing layer.

\begin{figure}[ht]  
\begin{center}
\resizebox{0.9\columnwidth}{!}
{\includegraphics{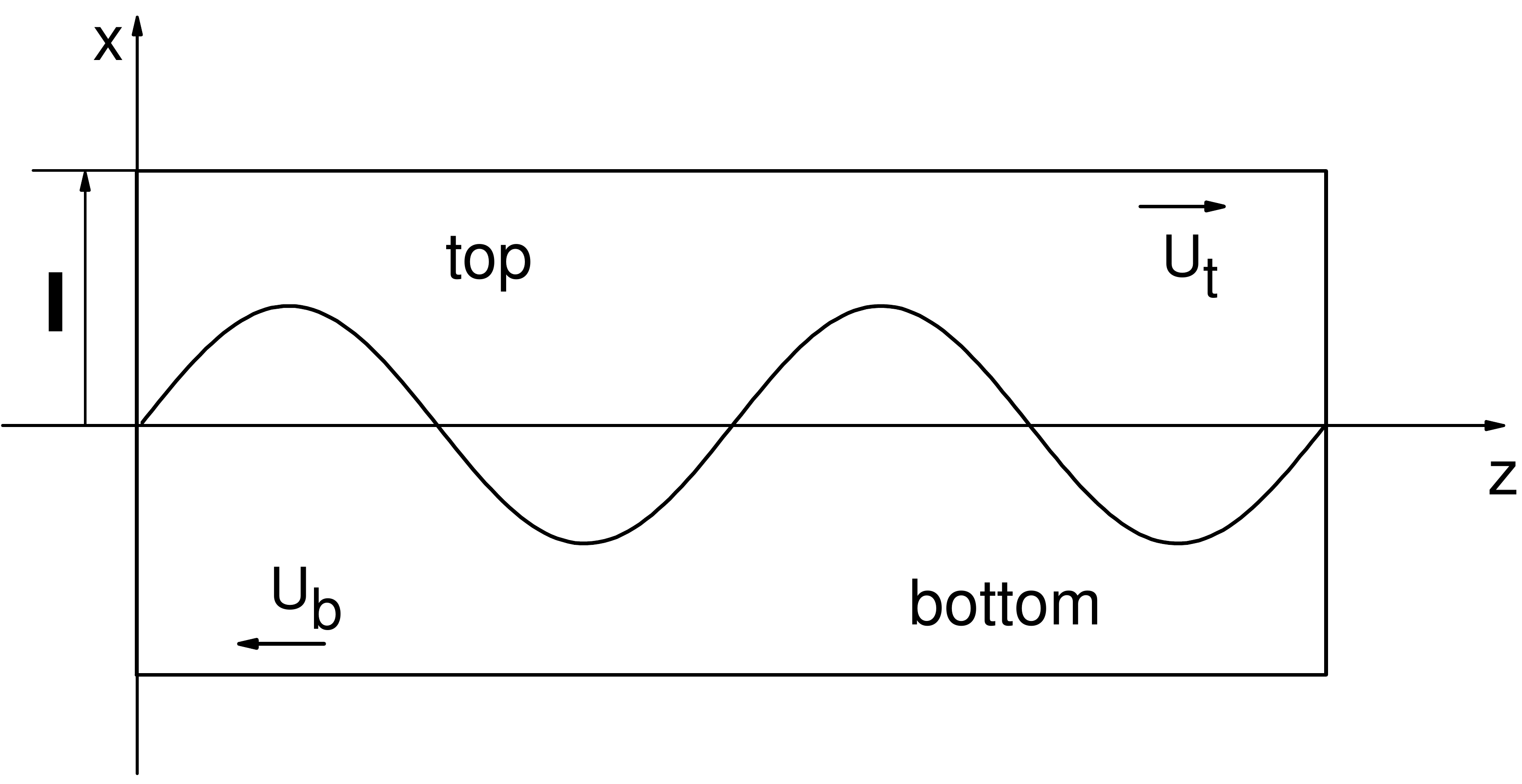}}
\caption{The profile of the top and bottom fluids layers with a dividing
surface wave on it.
The top external fluid moves with velocity $U_t$ along the $z$-direction,
while the bottom fluid
moves with velocity $U_b$ along $-z$-direction. The slip is unconstrained
at the dividing surface.
}
\label{profile}
\end{center}
\end{figure}

Following ref. \cite{FUNADA}
we idealize the problem and assume an initial state
where the shear is localized at the dividing plane between the top (t)
half and
the bottom (b) half of the fluids, in order that we can use the
potential flow description
in the top and bottom parts of the fluid, see Fig. \ref{profile}.
We assume that the fluid in the top and bottom
parts are allowed to slip at the top and bottom boundaries as
well as at the dividing surface between them. We will reference these
as unconstrained slip-conditions.
The initial flow velocity is assumed to be uniform in the
two layers, so that for the top layer
$\vec{v}^t=(0,U_t)$ for $0<x<l$ and
for the bottom layer $\vec{v}^b=(0,U_b)$ for $-l<x<0$
initially. This means that initially the amplitude of the
wave-like instability is extremely small,
and we are looking for the conditions to have a
growing amplitude for this instability.
For the sake of analytic model we assume that the density is constant.
Numerical studies \cite{CSA11} show that this constraint can be relaxed.

Just as in ref. \cite{CSA11} we will use an
average energy and mass density for our estimates. Since
the effective particle density is constant the continuity equation of
the flow velocity, $\vec{v}$, will become:
\begin{equation}
\nabla\cdot\vec{v}=0.
\label{C.E}
\end{equation}

For the top and bottom parts of the fluid we assume small velocities and
neglect velocity gradients and we assume that the rotation of
the flow velocity is ${\nabla} \times \vec{v}=0$. Under such conditions
we can describe
the flow as a potential flow, i.e., $\vec{v}=\nabla \phi$, where
$\phi$ is the velocity potential, $\phi\equiv\phi(t,x,z)$.
The continuity equation is
$\triangle\phi\equiv\nabla^2\phi=0$, applied this
for the top and bottom layers gives:
\ba
\nabla^2 \phi_t=0 \ &{\rm for}& \ 0<x<l,
\label{pot.phi.u}\\
\nabla^2 \phi_b=0 \ &{\rm for}& \ -l<x<0.
\label{pot.phi.d}
\ea

We assume that due to the raising instability the
initially plane interface will experience a perturbation
and will deviate from the $x=0$ plane. The height
of the deviation from the $x=0$ plane is denoted by $h=h(t,z)$, and
it is taken as wave-like perturbation in
the $z$ direction with wave-number $k$. We also allow the amplitude at a
given coordinate to change in time. The
most general form for such a wave-like interface would be
\begin{equation}
h(t,z)=A_0{\rm e}^{(\sigma t+ikz)},
\label{ele}
\end{equation}
where, $\sigma$ is the complex growth-rate,
and $A_0$ is a complex amplitude.

Consequently, the fluid on the top and bottom sides next to the
dividing surface will have vertical velocity components:
\ba
v_x^{t,b}=\frac{d h}{dt}&=&\frac{\partial h}{\partial t}+
U_{t,b}\frac{\partial h}{\partial z},
\label{B.C.}
\ea
where $U_t$ and $U_b$ are the flow velocities of the fluid at the
top and bottom of the bounding surfaces, respectively.
These are assumed to be the average velocity of the fluids at the surface
neglecting the horizontal velocity fluctuations arising from wave formation.
Initially these velocity fluctuations are small, and our aim is to study
the initial development of KHI. At the $h(t,z)$
dividing surface we also assume
unconstrained slip conditions of the two inversely flowing fluid slabs.

The boundary conditions for the external border of the profile are:
\ba
v_x^{t,b}=\frac{\partial \phi_{t,b}}{\partial x}=0  \ {\rm at} \ x=\pm  l,
\label{BCbor.u}
\ea
Initially (at time $t=0$), one would also have to satisfy:
\ba
v_z^{t,b}(t=0, x=\pm l)=U_{t,b}
\label{U_t_ini}
\ea

At each time moment the potential $\phi_t$  and $\phi_b$ is satisfying
Eq. (\ref{pot.phi.u},\ref{pot.phi.d},\ref{B.C.},\ref{BCbor.u})
for the top and bottom sides respectively:
\ba
\frac{\partial^2 \phi_{t,b}}{\partial x^2}&+&
\frac{\partial^2 \phi_{t,b} }{\partial z^2}=0
\nonumber\\
v_x^{t,b}=\frac{d h}{dt} = \frac{\partial h}{\partial t} &+&
U_{t,b}\frac{\partial h}{\partial z}\ \ \ {\rm at} \ x=0,\\
v_x^{t,b}=\frac{\partial \phi_{t,b}}{\partial x}&=&0\ \ \ \
{\rm at} \ x=\pm l.\nonumber
\ea
Assuming for the interface, $h(t,z)$, in Eq. (\ref{ele}), a wave-like
perturbation, which is symmetric in $\pm x$ and exponentially decreasing
away from the surface, the solution can be searched in the form:
\ba
\phi_{t,b}=A_{t,b} \ \cosh[k(x-l)]{\rm e}^{(\sigma t+ikz)}+z U_{t,b}\ ,
\label{sol.u}
\ea
where $A_t$, $A_b$ and $A_0$ are the complex amplitudes, $\sigma$
is the growth rate and $k$ is the wave number.
From the kinematic conditions on the dividing layer Eq. (\ref{B.C.}),
we  get
the following equations at the dividing surface:
\ba
(\sigma+ikU_{t,b})A_0&=&\mp k A_{t,b} \ \sinh(kl)
\label{cond}
\ea

The pressure ($p$), viscosity ($\eta$) and surface tension ($\gamma$)
balance at the interface writes as:
\begin{equation}
-p_t+2\eta\frac{\partial v_x^t}{\partial x}-(-p_b+2\eta
\frac{\partial v_x^b}{\partial x})=-\gamma\frac{\partial^2h}{\partial z^2},
\label{stress}
\end{equation}
The surface energy and consequently the surface tension of the
dividing layer will be approximated later. As it was already
emphasized in the introductory paragraphs,
although the top (t) and bottom (b) sides are of the same nuclear
matter, the velocity jump or the sharp velocity change contribute
to additional surface energy due to the
large shear at the interface exhibiting extra energy or to a smaller
extent by the momentum dependance of nuclear interaction potential.

Since we have unconstrained slip conditions on the
dividing surface between the top
an bottom layer,
$p_t$ and $p_b$ can be written by the classical
equation of motion without the viscous term as:
\ba
\rho(\frac{\partial v_z^{t,b}}{\partial t}+
U_{t,b}\frac{\partial v_z^{t,b}}{\partial z})=
-\frac{\partial p_{t,b}}{\partial z},
\ea
Then, first we apply $\nabla_z$ on both sides of the
equation and substitute equation of continuity,
$\partial_zv_z=-\partial_xv_x$, into it:
\ba
\rho(\frac{\partial^2v_x^{t,b}}{\partial t \partial x}+
U_{t,b}\frac{\partial^2 v_x^{t,b}}{\partial x\partial z})=
\frac{\partial^2 p_{t,b}}{\partial z^2}.
\label{p_t}
\ea
Here $\rho$ is the effective mass density of the QGP, we use
$\rho=10 \ {\rm GeV/fm^3c^2}$ \cite{CSA11} in our work.
In order to substitute the above equations into Eq. (\ref{stress}),
we consider the second order derivative of Eq. (\ref{stress})
as a function of $z$,
and substitute Eq. (\ref{p_t}) into it.
Thus the pressure, viscosity and surface tension
balance will be written in the following form:
\ba
-\rho(\frac{\partial^2 v_x^t}{\partial t \partial x}
+U_t\frac{\partial^2 v_x^t}{\partial x\partial z})
+2\eta\frac{\partial^3 v_x^t}{\partial x \partial z^2}+
\nonumber\\
\rho(\frac{\partial^2 v_x^b}{\partial t \partial x}
+U_b\frac{\partial^2 v_x^b}{\partial x\partial z})
-2\eta\frac{\partial^3 v_x^b}{\partial x \partial z^2}=
-\gamma\frac{\partial^4h}{\partial z^4}.
\label{sigmalcalculation}
\ea

By inserting the velocity derived from Eq. (\ref{sol.u}) and
the  considered interface profile, Eq. (\ref{ele}), into the above equation,
and expressing the top and bottom amplitudes, $A_{t,b}$
from Eq. (\ref{cond}), after
simplifying all over with $A_0$, and putting the condition $x=0$,
we obtain an equation for $\sigma$ and $k$:
\begin{eqnarray}
&[\rho&(\sigma+ikU_t)^2+2\eta k^2(\sigma+ikU_t)]\coth(kl) \nonumber\\
&+&[\rho(\sigma+ikU_b)^2+2\eta k^2(\sigma+ikU_b)]\coth(kl) \label{sigmaEq}\\
&+&\gamma k^3=0. \nonumber
\end{eqnarray}
Considering this as an equation for $\sigma$,
one can write it in a simplified form as
\ba
A\sigma^2+2B\sigma+C=0,
\label{dis-rel}
\ea
where the coefficients, $A,B,C$ are defined as:
\begin{eqnarray}
A&=&2\rho \ \coth(kl),\nonumber\\
B &=& 2k^2 \eta \, \coth(kl)+i k\rho\  (U_b{+}U_t) \coth(kl)\nonumber\\
&=& B_R{+}iB_I,\label{br}\\
C&=&-k^2\rho \ \coth(kl)(U_t^2+U_b^2)+\gamma k^3\nonumber\\
&+&2ik^3\eta \ \coth(kl)(U_t+U_b)=C_R+iC_I.\nonumber
\end{eqnarray}
The solution is
\begin{eqnarray}
\sigma&=&-\frac{B}{A}\pm\sqrt{\frac{B^2}{A^2}-\frac{C}{A}} \nonumber\\
\rightarrow \sigma_R+i\sigma_I&=&-\frac{B_R+iB_I}{A}\pm\frac{\sqrt{D}}{A},
\label{cond1}
\end{eqnarray}
where $D=D_R+iD_I$ and
\begin{eqnarray}
D_R&=&k^2\rho^2 \ \coth^2(kl)(U_t-U_b)^2
\nonumber\\
&+&4\eta^2k^4\coth^2(kl)-2\rho \ \coth(kl)\gamma k^3,\label{dr}\\
D_I&=&0,\nonumber
\end{eqnarray}
thus the real part and the imaginary part can be expressed as:
\be
\sigma_R=\frac{-B_R\pm\sqrt{D_R}}{A},\ \ \ \ \sigma_I=-\frac{B_I}{A} .
\label{sigmaDI=0}
\ee

\begin{figure}[h]  
\begin{center}
\resizebox{0.9\columnwidth}{!}
{\includegraphics{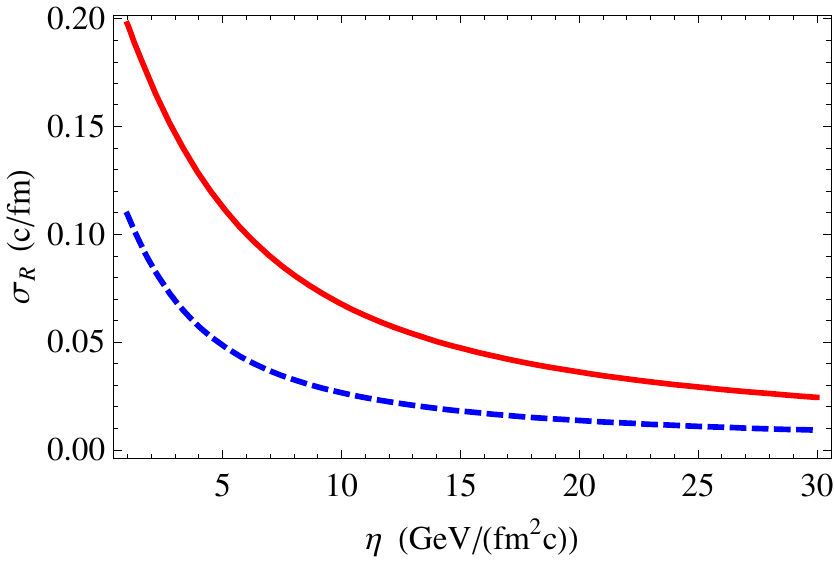}}
\caption{(color online)
The real part of the growth rate, $\sigma_R$, is shown as function of the
viscosity $\eta$. The full (red) line is for the surface tension
$\gamma=0.4  \ {\rm GeV/fm^2}$ and the dashed (blue) line is for
$\gamma=3.5 \ {\rm GeV/fm^2}$. The wave number, $k$, is taken to be
$k = 0.6 \ {\rm fm^{-1}}$ and the effective mass density is
$\rho = 10 \ {\rm Gev/fm^2 \ c^2}$. The growth rate decreases
when the viscosity increases suggesting that the KHI
grows weaker for a more viscous fluid.}
\label{diff-gamma}
\end{center}
\end{figure}
In heavy ion collisions, the matter will expand after the collision, and
in fact there is no external boundary (top and bottom) of the fluid
shown in Fig. \ref{profile}. If we assume $l\rightarrow\infty$,
the above equations can be simplified as:
\ba
\sigma_R &=& -\frac{k^2\eta}{\rho}\pm
\sqrt{ \frac{k^4 \eta^2}{\rho^2} +
\frac{k^2(U_t-U_b)^2}{4} -
\frac{\gamma k^3}{2\rho}   } \, ,
\label{SigmaR-l-infinite}  \\
\sigma_I &=& -\frac{k(U_t+U_b)}{2} \ .
\ea

\begin{figure}[ht]  
\vskip 2mm
\begin{center}
\resizebox{0.9\columnwidth}{!}
{\includegraphics{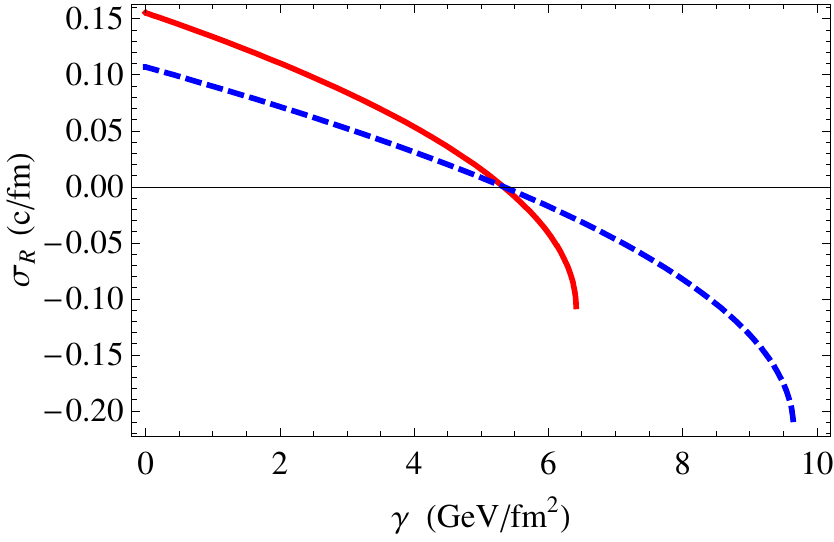}}
\caption{(color online)
The real part of the growth rate, $\sigma_R$, as a
function of the surface tension
$\gamma$ with different values of viscosity $\eta$. The full (red)
line represents
$\eta=3 \ {\rm GeV/fm^2} $ and the dashed (blue) line represents
$\eta=6 \ {\rm GeV/fm^2} $. The wave number $k$ is taken as
$0.6 \ {\rm fm^{-1}}$ and the effective mass density
$\rho$ is $10 \ {\rm Gev/fm^2 \ c^2}$. As we can see in the figure,
the two curves cross each other at $\sigma_R=0$, which is around
$\gamma=5.3 \ {\rm GeV/fm^2}$, and then the growth rate becomes negative.
With bigger surface tension the KHI effect is less probable to appear.}
\label{diff-eta}
\end{center}
\end{figure}
In Eq. (\ref{SigmaR-l-infinite}),
for the typical parameters of a peripheral heavy ion collision,
the real part of the growth rate,
$\sigma_R$, is dominantly dependent on the viscosity $\eta$,
namely the first term and the first term in the square root.
In our expanding system the dominant wave number of KHI is changing with
time.

\section{Results}

 According to the CFD observations \cite{CSA11},
initially we have a small
wave formation with $k\approx1\ {\rm fm}^{-1}$, but with time and expansion,
the possible largest wave length takes over with
$k\approx0.6 \ {\rm fm}^{-1}$,
which decreases further with the expansion of the system.  
By assuming $|U_t-U_b|=0.8$ c,
we can obtain the growth rate dependence of the viscosity, $\eta$,
and the surface
tension, $\gamma$, which are shown in Figs. \ref{diff-gamma}
and \ref{diff-eta}.

Similarly to the viscosity the effective surface energy also influences
the growth rate of KHI. As expected larger surface tension or surface
energy damps the growth of KHI. Beyond a critical surface energy
(in our model at $\gamma_{crit} \approx 5.3$ GeV/fm$^2$) the surface
tension will lead to a decrease in the KHI. Interestingly this threshold value
is independent of the viscosity. This is part of the general
feature that the behavior of the zero-growth ($\sigma_R=0$) curve
is independent of the value of the viscosity in this model.
The growth rate and damping rate are of course depend on the
viscosity.

The condition to have a growing instability is to have a solution with
$\sigma_R >0$.  Taking into account that $D$  is a
real number ($D_I=0$, \ $D_R>0$),  and $B_R>0$,  from (\ref{sigmaDI=0})
it follows that in order to have a positive growth rate, ($\sigma_R>0$),
one has to satisfy the condition:
\begin{equation}
\sqrt{D_R} \ > \ B_R \ .
\end{equation}
Thus, using (\ref{br}) and (\ref{dr}) we get the condition for
positive growth:
\ba
V^2>\frac{2\gamma k}{\rho \ \coth(kl)}\ ,
\label{critical-v}
\ea
where $V\equiv U_t-U_b$.

The above condition will limit the region of the ($V,k$) parameter space
where the KHI can evolve. One should also keep in mind the results obtained in \cite{CSA11}, regarding the
acceptable wave numbers, $k$, for the considered wave-like instability. Definitely there is a lower cutoff ($k_{min}$)
governed by the beam-directed longitudinal length of the flow, $l_z$:
\begin{equation}
k_{min}=\frac{2 \pi}{l_z}
\end{equation}
For the $b=0.5 b_{max}$ and $b=0.7 b_{max}$ impact parameter
values the calculations in
\cite{CSA11} leads to $k_{min}=0.598 \ {\rm fm}^{-1}$
and $k_{min}= 0.479 \ {\rm fm}^{-1}$ values, respectively.
There is also an upper limit for the wave-numbers, $k_{max}$
governed by the Kolmogorov length scale, $\lambda_K$:
\begin{equation}
k_{max}=\frac{2 \pi}{\lambda_K}
\end{equation}
According to \cite{CSA11} this characteristic length-scale is
estimated for the above given impact parameters as:
$\lambda_K \approx 3.5 \ {\rm fm}$
and $\lambda_K \approx 2.5 \ {\rm fm}$,
leading to $k_{max}=1.79 \ {\rm fm}^{-1}$ and
$k_{max}=2.51 \ {\rm fm}^{-1}$ values, respectively.

For the peripheral Pb+Pb collisions, the radius of Pb is $R=7$ fm,
thus $b_{max}=14$ fm.
In order to get the parameter space where the KHI will
growth let us estimate now the value of the surface tension.
As it has been discussed in the introductory part
this surface energy comes from the energy excess of the
unbalanced energy flow in the two layers.
Although a theory based on kinetic considerations would
capture more from the involved physics,
here we just consider a simple approach based on the
energy balance.  The reason for doing this is that less
number of phenomenological parameters are needed.

\begin{figure}[!htb]
\centering
\resizebox{0.9\columnwidth}{!}
{\includegraphics{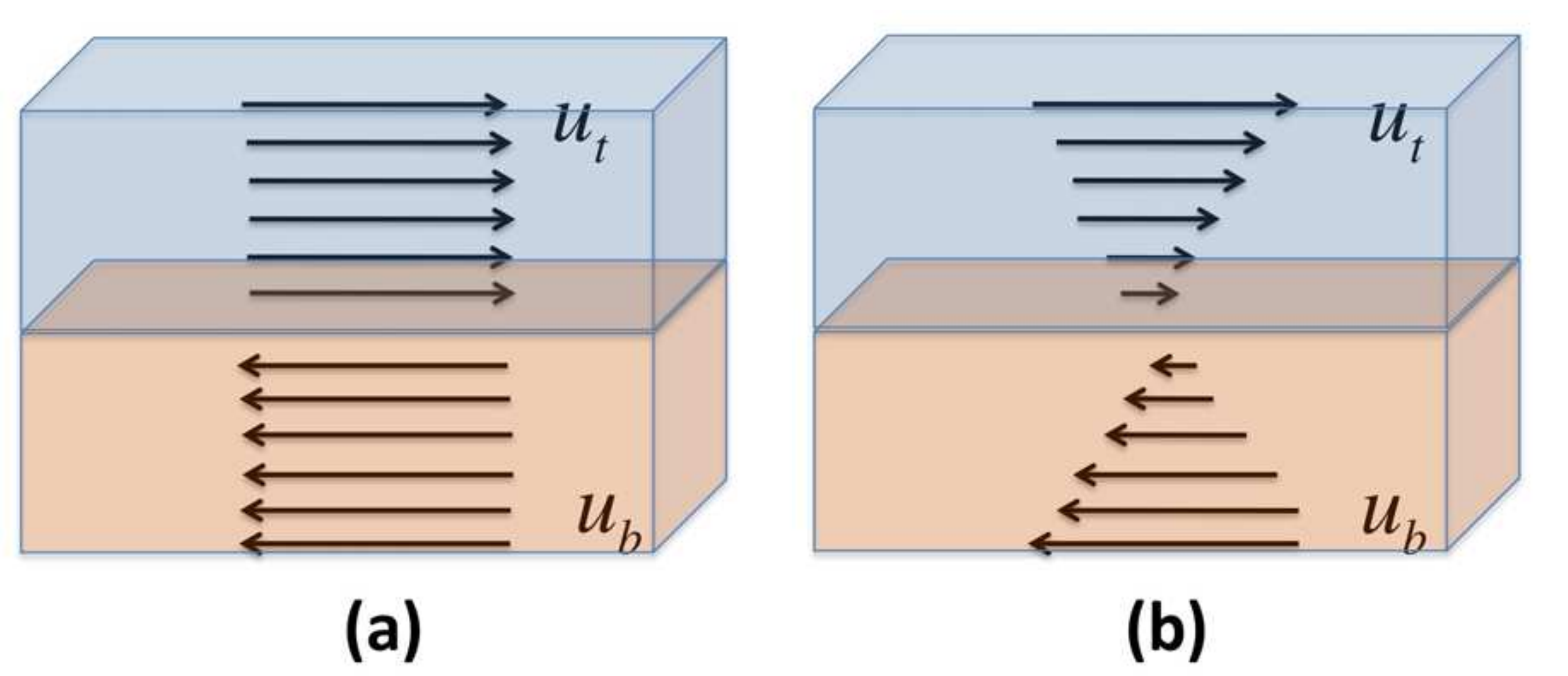}}
\caption{Two different velocity profiles: (a) is the profile used
in our present work, it has two distinct layers with two constant
velocity $U_t$ and $U_b$, while (b) has a flow transition from
the two layers, and at the diving surface the velocity is smallest.}
\label{surface-tension}
\end{figure}

The flow assumed in the present work has a perpendicular
velocity profile illustrated in Fig. \ref{surface-tension}a.
This means that a smooth velocity profile (Fig. \ref{surface-tension}b),
characterizing a stable and
balanced viscous flow is not formed. In the case
illustrated in Fig. \ref{surface-tension}a one would assume
that there are two distinct layers flowing with
velocities $U_t$ and $U_b$.  For the balanced flow illustrated
in Fig. \ref{surface-tension}b, one would observe
a smooth flow velocity transition from $U_t$ to
$U_b$. It is obvious that in the laboratory frame,
this later flow has less kinetic energy in the $z$ direction than the
previous one. The difference between the two kinetic
energies can be accounted as the energy surplus of the
dividing layer. If we denote the contact surface
between the flows in the top and bottom layers by $S$,
the surface tension could be estimated as

\ba
\gamma = \frac{E_{kz}^a-E_{kz}^b}{S},
\label{st}
\ea
where $E_{kz}^{a,b}$ denotes the kinetic energy
of the flow in the $z$ direction for the profile illustrated in Fig.
\ref{surface-tension}a and \ref{surface-tension}b, respectively.
The total relativistic kinetic energy of the
system, $E_k$ in the laboratory frame is
\ba
E_k=2M_{Pb}c^2\left (\frac{1}{\sqrt{1-\frac{V^2}{c^2}}}-1 \right ),
\label{energy-diff}
\ea
where $V=U_t-U_b$ is the relative speed of the
two projectiles, and $M_{Pb}$ is the mass of the
collided $Pb$ ions. Assuming that the participating
zone in the collision has a surface $q\times \pi R^2$,
(the overlapping regions are only  a $q$ part of
the possible ones) and the kinetic energy of the participating particles
in this zone is distributed equally in all the
directions of the space, a rough approximations for $E_{kz}^a$ would be:
$E_{kz}^a=q \frac{E_k}{3}$. On the other hand,
for the flow illustrated in Fig. \ref{surface-tension}b,
due to the balanced velocity profile, a part of this
kinetic energy has to be dissipated, and assuming a linear
velocity profile one gets:
$E_{kz}^b=1/2 \ E_{kz}^a$.
The above arguments lead us to a first approximation
of the surface tension value:
\ba
\gamma =
\frac{q}{3} \frac{M_{Pb}c^2}{S}
\left (\frac{1}{\sqrt{1-\frac{V^2}{c^2}}}-1 \right ).
\label{stf}
\ea

Assuming $q \approx 0.5$ and estimating the surface of the dividing
layer, $S$, from \cite{CSA11}, one gets the values of $\gamma$
for different impact parameter values.

\begin{figure}[ht]  
\begin{center}
\resizebox{0.95\columnwidth}{!}
{\includegraphics{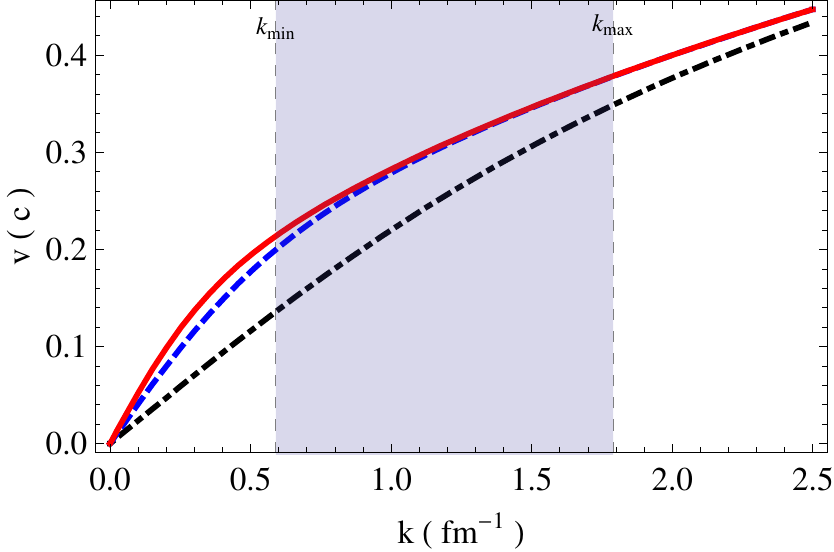}}
\caption{(color online) The critical velocity $V\equiv U_t-U_b$ as
function of the wave number, at the condition of vanishing growth
rate, $\sigma_R=0$. The red full line, blue dashed line and black
dot-dashed line are for impact parameters $b=0.5, 0.7, 0.9 \  b_{max}$,
respectively.
On the graph we also illustrated the two natural boundaries
$k_{min}$ and $k_{max}$ for $b=0.5 \ b_{max}$. The KHI will evolve
thus above the critical velocity curves and between these two limits.
For increasing impact parameters, the instability is less able to
grow and the system tends to be stable.
}
\label{vk}
\end{center}
\end{figure}

The surface tension is estimated to be $\gamma=0.4 \ {\rm GeV/fm^2}$
from Eq. (\ref{stf}). This value is used in the following examples.
The critical velocity Eq. (\ref{critical-v}) for different impact
parameters is shown in Fig. \ref{vk}.
These curves show the border of instability of the growth rate, $\sigma_R$.
The curves divide the space into two areas, the upper side above the
curve is the region where the instability grows and the
area below the critical velocity curve is where the instability does not grow.
The KHI development region is also limited by the $k_{min}$ and $k_{max}$
values as drawn in figure Fig. \ref{vk}.

\begin{figure}[!htb]
\centering
\resizebox{0.9\columnwidth}{!}
{\includegraphics{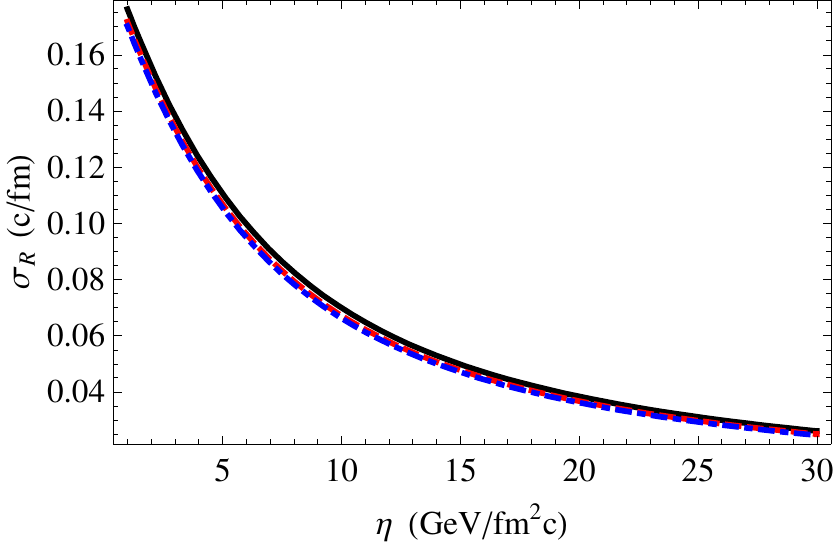}}
\caption{(color online)
The growth rate $\sigma_R$ as a function of the
viscosity $\eta$ at different values of  $l$, black line is for
$l=0.1$ fm, the red dashed line is for $l=2$ fm and the blue dot-dashed line
is for  $l= \infty$.  The wave number is $k=0.6 \ {\rm fm^{-1}}$, the surface
tension is $\gamma=0.4 \ {\rm Gev/fm^2}$,
 the relative velocity is $V = 0.8$ c
and $\rho=10 \ {\rm Gev/fm^2 \ c^2}$. The growth rate  depends
weakly on $l$, while it depends significantly on the viscosity, increasing
strongly for small viscosity values.}
\label{diff-l}
\end{figure}

The above consideration is for $\sigma_R=0$, however,
this does not show the $\eta$-dependence of the growth. In order to see
how the instability depends on the viscosity, $\eta$, we
can cast Eq. (\ref{sigmaDI=0}) into the form:
\be
\sigma_R = \frac{k^2\eta}{\rho}
\left[ -1 \pm \sqrt{ 1+ \frac{\rho }{\eta^2}
\left( V^2 \rho - \frac{\gamma k}{\coth(kl)}\right)\,}\, \right] \ .
\label{SigmaR-l}
\ee
This suggests that with our characteristic parameters the
dependence on the thickness of the fluid layer, $l$, is weak
as  shown in Fig. \ref{diff-l}.
\bigskip

\section{Conclusions}

In classical gravitational water waves the wave formation and wave
speed depends strongly on the depth of the water, i.e. the layer
thickness, $l$.  In a heavy ion collisions the role of the
layer thickness is different. The material properties of the
top and bottom layers are not different, these are separated
from another by the relatively thin layer of large shear.
Still the occurrence of KHI in such conditions is not uncommon
as it is frequently
observed, as turbulence during airplane flights, or it is even
visible if the air has high humidity and the condensation makes
the KHI visible.

In peripheral heavy ion collisions the layer thickness is given in the
initial state, but there is no solid boundary and the system expands
in all directions. Thus, for this physical situation the large
or infinite layer thickness is more relevant in this model, even if
the initial layer thickness is finite and usually smaller than the
longitudinal size of the initial state.

Large viscosity or the corresponding low Reynolds number
prevent the development of turbulence and KHI, so that these
phenomena appear only above a critical Reynolds number. This
critical Reynolds number depends on the flow configuration, so it
is separately analysed for the KHI also, see ref. \cite{CSA11}.
The present study confirms that the dependence of the
growth rate on the viscosity reflects the usual tendency that
instability and turbulence increases with smaller viscosity.

When the KHI develops between two fluids (e.g. air/water or air/oil)
the large surface tension difference at the interface damps the development
of the instability, this is well known for sailors for centuries. If
KHI develops inside one fluid, like in air or in quark gluon fluid, there is
no surface tension in the classical sense, but the layer with large shear
has extra energy, and it leads to an effective surface tension, which
hinders the development of KHI.

We presented a strongly idealized analytic model for the development
of the Kelvin-Helmholtz Instability in ultra-relativistic heavy ion
reactions. We compressed the shear zone into a central infinitesimal
layer, following the idea of ref. \cite{FUNADA}, and assumed that
the remaining flow can be approximated as potential flow. The idealized
dividing layer was attributed a surface energy and unconstrained slip
between the top and bottom fluid layers. It is interesting that in this
model the KHI is developing under similar conditions,
as in numerical high resolution relativistic fluid dynamical
calculations \cite{CSA11}. This model also shows that critical size
KHI may occur for low viscosity QGP.

{\em Acknowledgement} The work of Z.N. was supported by the exploratory "Ideas" research
grant: PCE-IDEI-0348/2011.

\end{document}